\documentclass[12pt]{iopart}
\usepackage[utf8]{inputenc}
\usepackage{graphicx}
\usepackage{hyperref}
\usepackage[dvipsnames]{xcolor}

\hypersetup{
    colorlinks=true,
    linkcolor=blue,
    citecolor=ForestGreen,
    filecolor=magenta,      
    urlcolor=cyan,
    pdftitle={Overleaf Example},
    pdfpagemode=FullScreen,
    }

\newcommand{\affilipp}{Max-Planck-Institute for Plasma Physics, Boltzmannstr.~2, 85748 Garching, Germany }
\newcommand{\affilepfl}{Ecole Polytechnique Federale de Lausanne, Swiss Plasma Center, CH-1015 Lausanne, Switzerland}
\newcommand{\affilsev}{Dept. of Atomic, Molecular and Nuclear Physics, University of Seville, Spain}
\newcommand{\affilccfe}{United Kingdom Atomic Energy Authority, Culham Science Centre, Abingdon, Oxon, OX14 3DB, United Kingdom of Great Britain and Northern Ireland}
\newcommand{\affilmit}{Plasma Science and Fusion Center, Massachusetts Institute of Technology, Cambridge, MA 02139, United States of America}
\newcommand{\affilbrus}{Laboratory for Plasma Physics LPP-ERM/KMS, B-1000 Brussels, Belgium}

\newcommand{\affiltuwien}{Institute of Applied Physics, TU Wien, Fusion@\"OAW, Wiedner Hauptstr. 8-10, 1040 Vienna, Austria}
\newcommand{\affilcie}{Laboratorio Nacional de Fusión, CIEMAT, Spain}
\newcommand{\affiljet}{See the author list of C. Maggi et al, 2024 Nucl. Fusion https://doi.org/10.1088/1741-4326/ad3e16}
\newcommand{\affilaug}{See the author list of H. Zohm et al, 2024 Nucl. Fusion https://doi.org/10.1088/1741-4326/ad249d}
\newcommand{\affiltcv}{See author list of B.P. Duval et al. 2024 \emph{Nucl. Fusion} \textbf{64} 112023}
\newcommand{\affilwpte}{See the author list of N. Vianello et al, ''Results from the last DD and DT JET campaigns in the framework of the EUROfusion Tokamak Exploitation Work Package Activity", 2026 submitted to Nuclear Fusion}

\begin{document}
\title{The physics of ELM-free regimes in EUROfusion tokamaks}

\author{M.G. Dunne$^1$, M. Faitsch$^1$, O. Sauter$^2$, E. Viezzer$^3$, B. Labit$^2$, A. Kappatou$^1$, D. Keeling$^4$, B. Vanovac$^5$, I. Balboa$^4$, \\
P. Bilkova$^5$, P. Bohm$^5$, D. Kos$^4$, J. Hobirk$^1$, E. Lerche$^{6,4}$, \\
P. Lomas$^4$, S. Menmuir$^4$, T. P\"utterich$^1$, L. Radovanovic$^7$, \\
S. Saarelma$^4$, S. Silburn$^4$, D. Silvagni$^1$, E.R. Solano$^8$, H.J. Sun$^4$, A. Tookey$^4$, The ASDEX Upgrade Team$^\dagger$, The TCV Team$^\ddagger$, The EUROfusion Tokamak Exploitation Team$^\flat$, and JET contributors$^\natural$}
\address{$^1$\affilipp}
\address{$^2$\affilepfl}
\address{$^3$\affilsev}
\address{$^4$\affilccfe}
\address{$^5$\affilmit}
\address{$^6$\affilbrus}
\address{$^7$\affiltuwien}
\address{$^8$\affilcie}
%devices
\address{$^\dagger$\affilaug}
\address{$^\ddagger$\affiltcv}
\address{$^\flat$\affilwpte}
\address{$^\natural$\affiljet}

\begin{abstract}
The development of operational scenarios without large Type-I ELMs is of utmost importance for the stable operation and longevity of future tokamaks. The EUROfusion tokamak exploitation program has therefore made the understanding of ELM-free regimes a major topic of exploration across all its contributing devices (ASDEX Upgrade, JET, MAST-Upgrade, TCV, and WEST). An integrated program to investigate a range of Type-I ELM-free regimes has been developed covering the enhanced D-alpha (EDA), magnetic perturbations (MP), negative triangularity (NT), quasi-continuous exhaust (QCE), quiescent H-mode (QH), the baseline small ELMs (SE), I-mode, and X-point radiator (XPR) regimes. This contribution focuses on the development and understanding of the NT and QCE regimes on ASDEX Upgrade, JET, and TCV. The importance of transport via ballooning modes in both regimes is highlighted, as well as the progress in developing access models based on ideal-MHD. In the case of the QCE, this can also be expressed as a minimum separatrix density, which corresponds well to experimentally measured separatrix densities. Particular focus is paid to the performance of the QCE in terms of the achieved pedestal top values, which, when appropriately normalised, do not differ significantly from ELMy H-mode plasmas. This, combined with the predicted minimum separatrix density for the 15~MA ITER baseline plasma, highlight the relevance of the QCE as a potential operational scenario for both ITER and future reactors.
\end{abstract}

%\ioptwocol

\section{Introduction}
The development of operational scenarios without large Type-I ELMs is of utmost importance for the stable operation and longevity of future tokamaks. To this end, one of the most important research topics at the moment is the search for robust ELM-free regimes with sufficient confinement to support a fusion reactor. The EUROfusion tokamak exploitation program has therefore made the understanding of ELM-free regimes a major topic of exploration across all its contributing devices (ASDEX Upgrade, JET, MAST-Upgrade, TCV, and WEST). 

The overarching aim of the research topic is directed at the physics understanding of large-ELM-free regimes, rather than the development of an integrated high performance, detached ELM free scenario. The integration of a high performance core with an ELM-free pedestal and relevant divertor physics is quite challenging in present-day machines owing to the typical requirement of high separatrix density operation, which imposes known limitations on plasma performance\cite{Dunne2017,Frassinetti2021,Sheikh2019}. Instead, the general approach of the topic is to investigate physics mechanisms in smaller, flexible devices, such as TCV and ASDEX Upgrade, to validate this understanding on a larger device (typically JET) and then use the validated physics mechanisms to make predictions for future devices such as SPARC and ITER. 

A wide range of scenarios are being investigated across the different machines, covering the enhanced D-alpha (EDA)\cite{Greenwald1999,Gil2025}, ELM suppression via magnetic perturbation (MP)\cite{Evans2004,Nazikian2015,Suttrop2017}, the I-mode\cite{Hubbard2016, Happel2017}, negative triangularity(NT)\cite{Marinoni2021,Nelson2022,Coda2021,Happel2022}, the quasi-continuous exhaust (QCE) regime\cite{Harrer2018,Labit2019,Faitsch2023,Faitsch2025}, quiescent H-mode (QH)\cite{Burrell2002,Suttrop2004,Suttrop2005,Solano2010,Chen2020}, the baseline small ELMs (SE)\cite{Garcia2022,delaLuna2024}, and X-point radiator (XPR) regimes\cite{Bernert2025}. The focus of this contribution is on the well-developed understanding of NT and QCE plasmas. A highlight of these efforts is the development and exploration of NT and QCE scenarios in JET, enabled via predictive modelling. A second highlight of the recent JET ELM-free experiments is the demonstration of both the QCE\cite{Faitsch2025} and XPR\cite{Bernert2025} regimes in DT plasmas.

The manuscript is structured as follows: an introduction to a simple conceptual model of ELM-avoidance is presented, along with the tools which are used for modelling. The progress in developing ELM-free negative triangularity scenarios on ASDEX Upgrade and JET, following detailed experiments at TCV will be mentioned. The paper will, however, focus more on recent progress in the QCE regime, describing the current access model and a comparison of the scenario with ELMing H-modes on ASDEX Upgrade, JET, and TCV.

\section{The physics picture of Type-I ELM avoidance}
\begin{figure}
\centering
\includegraphics[width=0.4\textwidth]{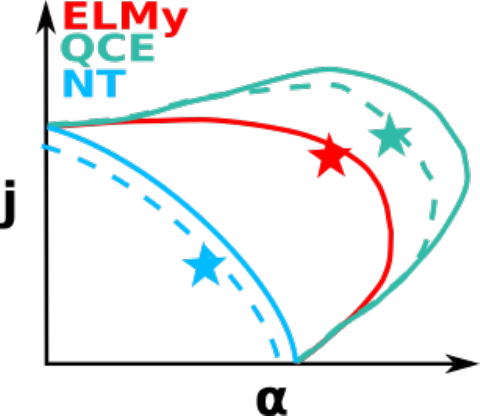}
    \caption{Schematic of Type-I ELM avoidance for QCE (green) and NT (blue) plasmas vs an ELMy H-mode (red) in a j-$\alpha$ diagram where j is the peak edge current density and $\alpha$ is the normalised pressure gradient. The stars indicate the operational point expected in an experiment. The region of stability lies in the lower left of the plot, and above and to the right of each curve is the unstable region.}
    \label{fig:eped}
\end{figure}

The occurrence of Type-I ELMs can be understood as an intersection of transport in the pedestal (often considered to be a KBM\cite{Snyder2009}) determining the average pedestal gradient and the onset of a global peeling-ballooning mode. This is sketched in figure \ref{fig:eped}, using the well-known j-$\alpha$ depiction for global pedestal stability; above and to the left of the lines are the unstable region, below and left shows the stable region, where pedestals are expected prior to large ELM crashes. Figure \ref{fig:eped} shows an example stability boundary and operational point for an ELMy pedestal as the red line, where the transport and MHD limits coincide\cite{Snyder2011,Dickinson2011,Wolfrum2015}. An experimental plasma is expected to exist on this red line, indicated by the star.

A large ELM crash can be avoided by tailoring both the transport and MHD limits. There are then two main categories of large-ELM avoidance: one applies to e.g. QCE\cite{Kalis_2024} and QH mode\cite{Burrell2002}, where the Type-I ELM stability limit is raised (e.g. via increased plasma shaping) and additional transport is created via an MHD mode. This corresponds to the green lines in figure \ref{fig:eped}. In this case, the solid line indicates the global MHD stability limit, and the dashed line to the (slightly lower) transport limit. This allows an ELMy H-mode-like pedestal gradient to be sustained, shown by the green star, without the presence of large ELMs. The second category reduces both stability limits significantly, to the point where the H-mode is no longer accessed, as is the case in NT plasmas, corresponding to the blue lines in figure \ref{fig:eped}; here, the star indicates the low expected pedestal gradient, which remains below the peeling-ballooning boundary.

The MHD limit and its dependencies are generally well understood; increasing the plasma shape, generally thought of as a combination of both elongation and triangularity\cite{Dunne2024}, as well as the global plasma $\beta$\cite{Dunne2017a} and q$_{95}$\cite{Snyder2019} tend to increase the stability limit. The main task of understanding the physics of ELM-free regimes is then one of understanding the transport mechanisms which regulate the pedestal gradient and the pedestal width. 

\subsection{Regions of the pedestal}
The description of pedestal structure and transport in terms of only the pedestal height and width does not capture the finer details of the various transport mechanisms which can be present in the pedestal region. Gyrokinetic studies\cite{Leppin2023} of the ASDEX Upgrade pedestal have shown a range of instabilities which can be present at a given time. With this in mind, we can consider three regions of the pedestal: the top, close to the knee where the pedestal meets the core plasma; the middle, where the steepest gradients are located; and the foot, close to the separatrix. 

QH-mode plasmas are understood in the MHD framework; they rely on a saturated large-scale kink/peeling mode to provide extra transport. Recent modelling of a transient QH-mode phase in ASDEX Upgrade\cite{Meier2023} has confirmed the physics picture, and highlighted the importance of low pedestal collisionality in sustaining the QH-mode in these plasmas. 

Previous studies\cite{Snyder2012} have posited the mechanism for ELM suppression via magnetic perturbations as being caused by the presence of an island at the pedestal top, blocking further expansion of the pedestal. An island has recently been observed in ELM-suppressed ASDEX Upgrade plasmas with magnetic perturbations\cite{Willensdorfer2024}.

Ballooning modes have been shown to play the dominant role in NT\cite{Nelson2022,Nelson2023} and QCE\cite{Radovanovic2022} plasmas. In the case of NT, H-mode avoidance has been linked with blocking access to the second stable region in s-$\alpha$ space\cite{Nelson2022}, particularly in the middle region of the pedestal, where second-stability access is most likely. Access to the QCE is linked to keeping access to second-stability open in the middle of the pedestal and an unstable ballooning mode localised at the pedestal foot\cite{Harrer2018,Griener2020,Kalis_2024}, independently of the structure of the rest of the pedestal (which may tend towards low-, medium-, or high-n modes).

\section{Second stability access: negative triangularity}
\begin{figure}
    \centering
    \includegraphics[width=0.5\textwidth]{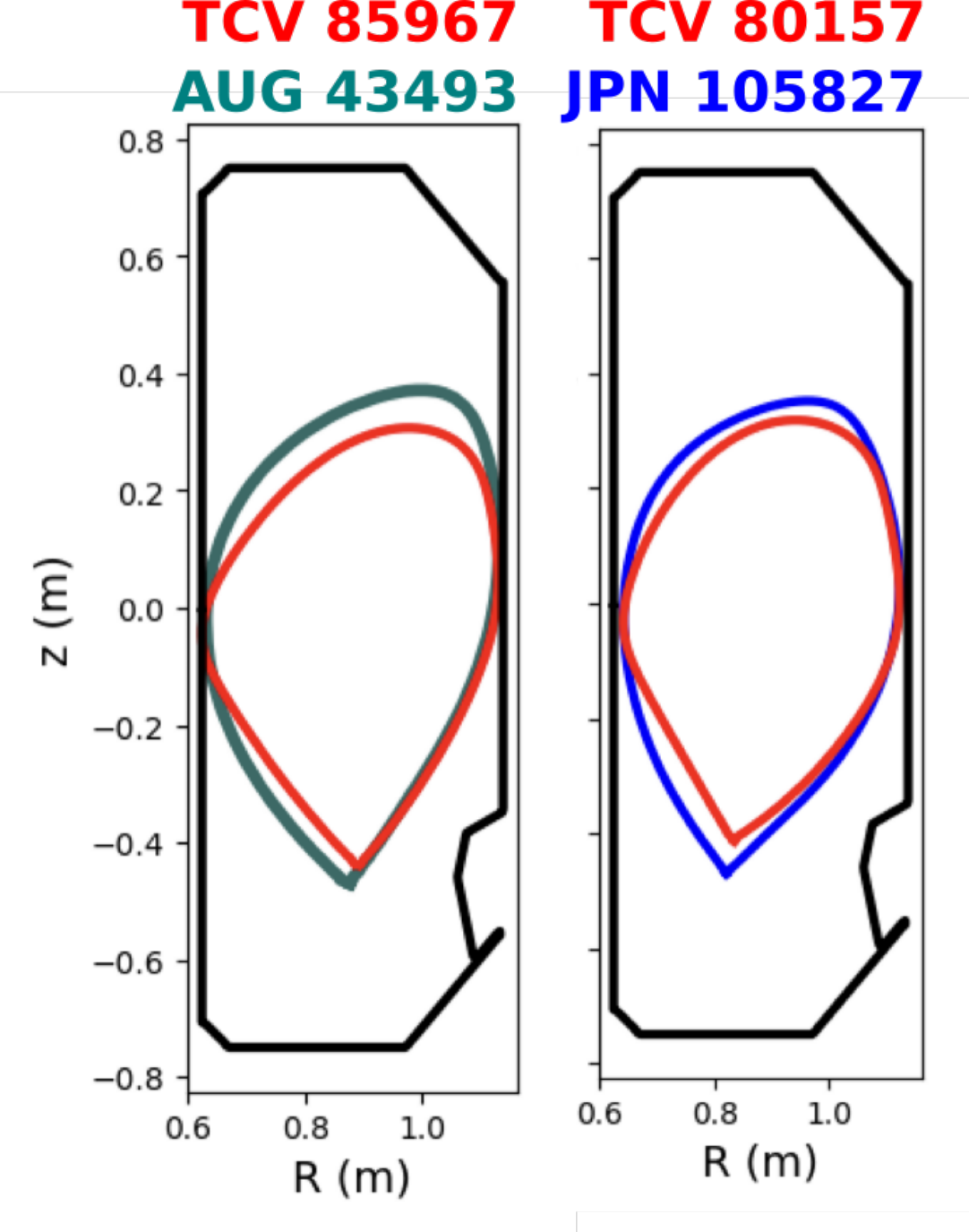}
    \caption{Left: AUG NT shape (green) and its TCV counterpart (red). Right: the JET NT shape (blue) and its TCV counterpart (red).}
    \label{fig:TCV_NT_shapes}
\end{figure}

The flexible shaping capabilities of TCV have helped drive the understanding of the separate elements of ELM-avoidance and confinement improvement in NT. Initial negative triangularity discharges at AUG showed no signs of ELM-avoidance or H-mode avoidance\cite{Happel2022,Vanovac_2024}. Since the original experiments the understanding of NT ELM avoidance\cite{Nelson2022,Nelson2023,Sauter2023} has developed, showing that blocking access to ballooning second stability avoids H-mode entry. These predictions indicated that minor changes to the AUG NT shape should be sufficient to block second stability access. In parallel, modelling also indicated that a negative triangularity shape which could be performed on the JET tokamak should also avoid H-mode access. Scaled versions of each of these shapes were developed on TCV and, combined with an NBI heating power ramp, verified H-mode avoidance in the range of relevant shapes. The shapes developed at TCV and their AUG and JET counterparts (scaled to TCV) are shown in figure \ref{fig:TCV_NT_shapes}.

\begin{figure}
    \centering
    \includegraphics[width=0.5\textwidth]{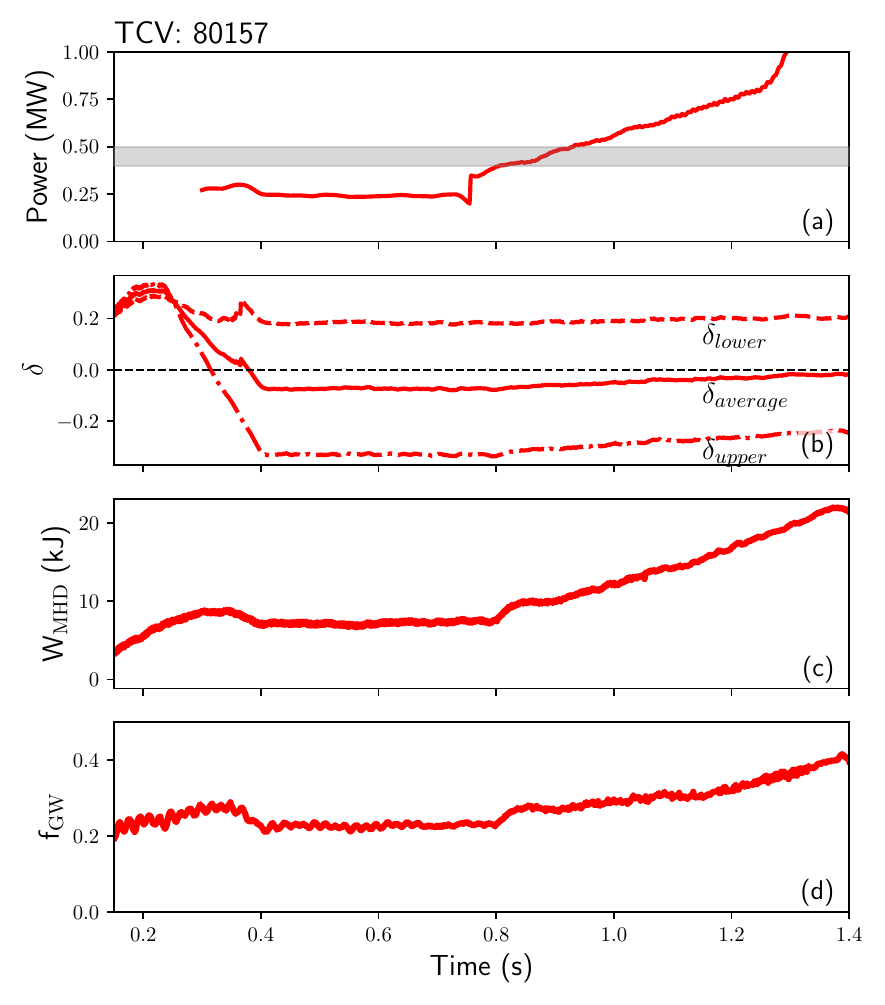}
    \caption{NBI heating power (a), upper (dash-dot), lower (dashed), and average (solid) triangularity (b), plasma stored energy (c), and Greenwald fraction (d) for the TCV companion plasma to the JET NT shape. Despite the significant injected power no LH transition can be seen.}
    \label{fig:TCV_JET_NT}
\end{figure}

The key feature of these experiments is the validation of the modelling-driven stepladder approach to the experiments. Predictions for the minimum shaping for ELM avoidance were made for JET and AUG, taking the restrictions of the PF coil-sets into account. Mock-up experiments were then performed on TCV, with an example shown in figure \ref{fig:TCV_JET_NT}. NBI heating power, shown in panel (a) was steadily increased well beyond the typical LH-transition power for similar plasmas in TCV at constant shaping (panel (b)). Panels (c) and (d) show the plasma stored energy and density increasing steadily with heating power, indicating no change of confinement regime. Similar experiments were performed for the AUG discharges, also showing robust H-mode avoidance. 

\subsection{H-mode avoidance via NT in AUG}
\begin{figure}
    \centering
    \includegraphics[width=0.5\textwidth]{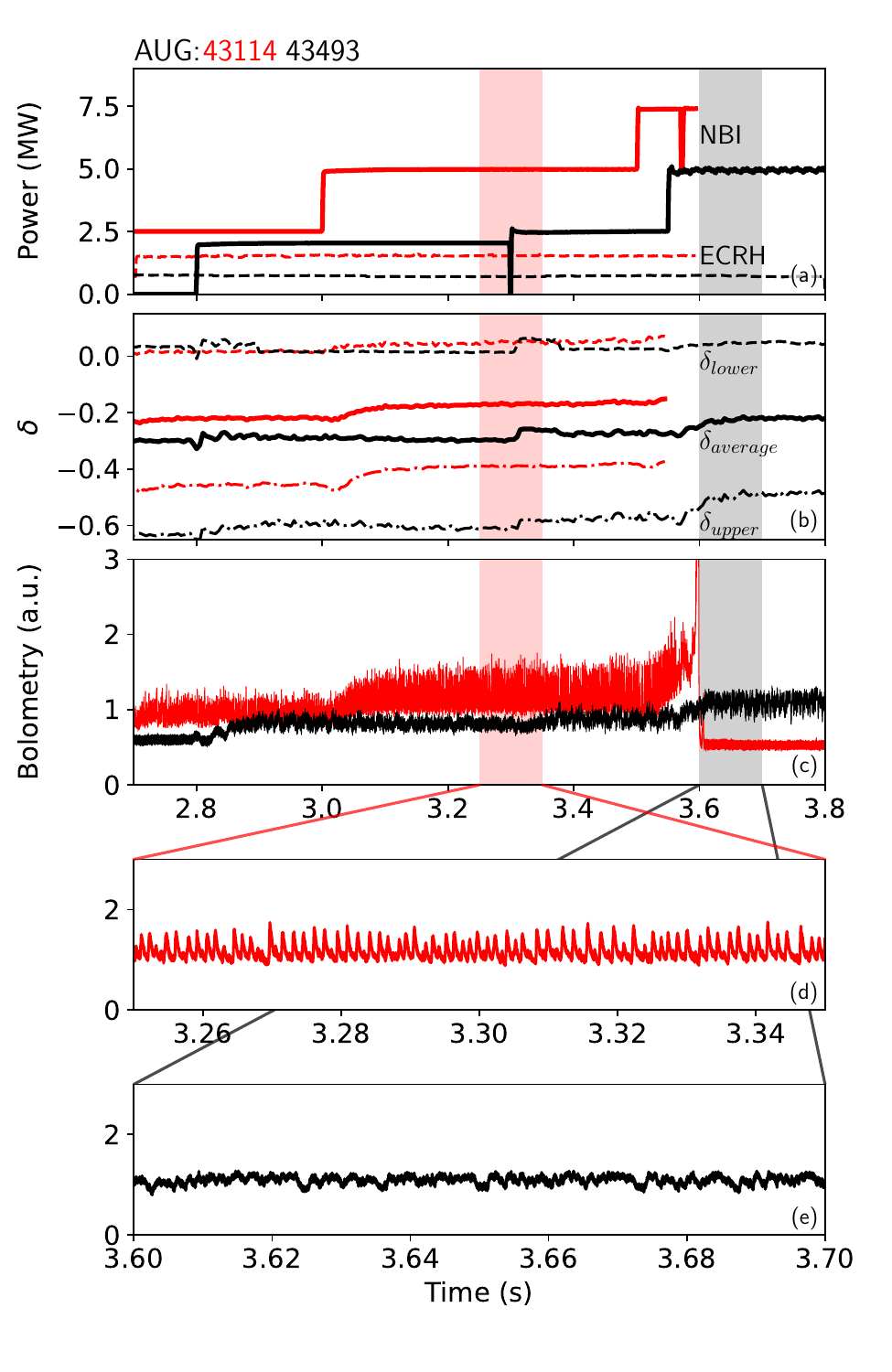}
    \caption{Timetraces of AUG discharges 43114 (red, ELMy) and 43493 (black, ELM-free) showing (a) heating power, (b) upper (dot-dashed), lower (dashed), and averaged (solid) triangularity, (c) the signal from a bolometry LOS acting as ELM monitor, (d) zoom into the ELM signal for the ELMy discharge, and (e) zoom onto the same signal for the ELM-free discharge during a phase of heating power comparable to the time window in panel (d).}
    \label{fig:AUG_NT_noelms}
\end{figure}

New negative-triangularity experiments, motivated by advances in prediction, were performed in ASDEX Upgrade during the 2025 campaign; the full details of the shape development, modelling, and plasma performance will be detailed in a separate publication\cite{Vanovac2026}. Here, we present two AUG discharges with small differences in the plasma shape; figure \ref{fig:AUG_NT_noelms} shows time-traces from a less strongly shaped plasma (red) and a more strongly shaped (i.e. less negative triangularity) plasma (black). The injected heating power (a) has comparable phases in both plasmas (at 3.1~s in the less shaped, and 3.7~s in the more strongly shaped plasma). The upper (dot-dashed), lower (dashed), and average (solid) triangularities (b) show the similarity in the lower shaping, with the difference in upper and average triangularity clearly visible. 
Panel (c) shows time-traces from a bolometry line-of-sight which indicates the ELM behaviour in both discharges. Panels (d) and (e) show a zoom onto the two comparably heated time windows. The less-shaped discharge (red, (d)) shows clear individual events, indicating ELM activity. The more strongly shaped discharge (black, (e)) shows no clear individual events; at most there is some dithery behaviour perhaps indicating that the discharge is close to an LH transition, but no ELMs are clear. 

\subsection{Demonstration of NT ELM-avoidance in JET}
A strenuous test of the H-mode avoidance theory was facilitated via the opportunity to develop and test a negative triangularity scenario on JET. Due to the high-risk nature of the experiments (both programatically and to the machine), only the final experimental day was dedicated to this program. Advance preparation, as mentioned above, predicted that the shaping which could be achieved in JET was sufficient for ELM avoidance. At the same time, further experiments in TCV based on this shape indicated that much more strongly shaped plasmas would be necessary for confinement improvement; this could not be tested due to limitations of the PF-coils at JET. As such, the goal of these experiments was ELM avoidance and not confinement studies.
\begin{figure}
    \centering
    \includegraphics[width=0.5\textwidth]{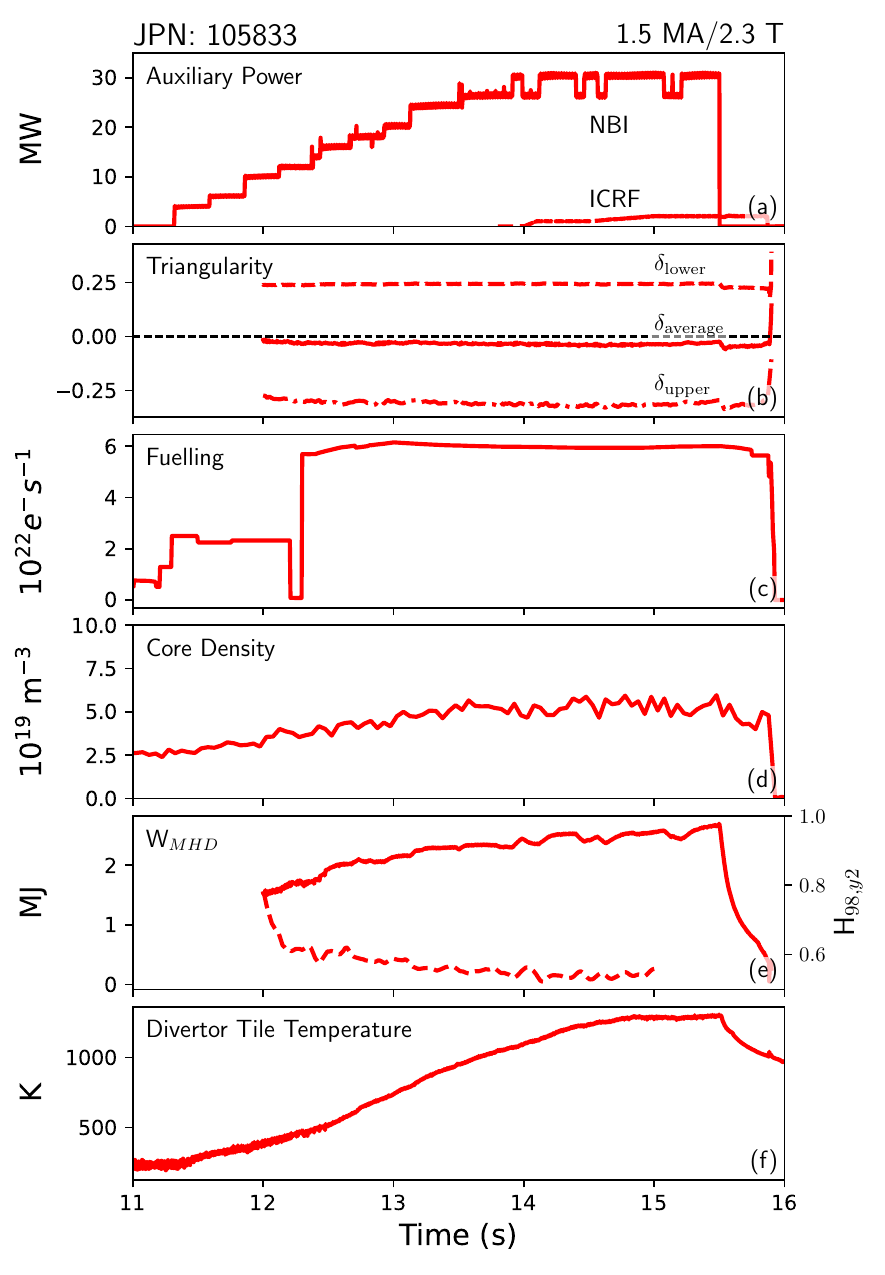}
    \caption{Time-traces of (a) injected heating power, (b) plasma density, (c) plasma stored energy (solid) and H$_{98,y2}$ (dashed), (d) and (e) ELM monitors.}
    \label{fig:jet_nt}
\end{figure}

Owing to the preparation of the experimental and engineering teams at JET the development and execution of an NT scenario at 1.5 MA could be realised within this single experimental day. While a more detailed analysis of the various discharges will be presented elsewhere\cite{Sauter2024,Sauter2026}, one highlight is presented here in figure \ref{fig:jet_nt} as time-traces of a high-power negative triangularity discharge. The injected heating power (a) reached a maximum of 32~MW, far beyond the expected LH transition power for a 1.5~MA/2.3~T discharge in JET (ca. 5-6~MW P$_\mathrm{loss}$, depending on the scaling using\cite{Solano2023}). The plasma density (b) and stored energy (c, solid line) display no sudden change in trajectory over the discharge, indicating no change of the confinement regime. One significant drawback of this discharge is seen in sub-figure (c) as the dashed line, showing the value of H$_{98,y2}$, indicating quite low normalised confinement; high confinement was not expected in this discharge, as outlined previously.

The ELM monitor signals in panels (d) and (e) show no ELM-like activity over the majority of the discharge. Filaments become visible only at the highest level of input heating power, and are reminiscent of the activity seen in the QCE discharges at JET, rather than type-I ELMs. A direct comparison between QCE and NT discharges at JET will be shown in section \ref{sec:qce_nt}. Work is still ongoing to characterise the transport across the entire plasma radius in the full set of JET-NT discharges. These experiments have, however, shown that robust ELM avoidance can be obtained in a large device, at substantial levels of heating power, and that the shaping required for this can be predicted in advance via only the engineering parameters (current, field, and plasma shape), building confidence in our ability to design NT ELM-avoidance scenarios in future devices.

\section{Separatrix ballooning modes: the quasi-continuous exhaust regime}
\begin{figure}
    \centering
    \includegraphics[width=0.5\textwidth]{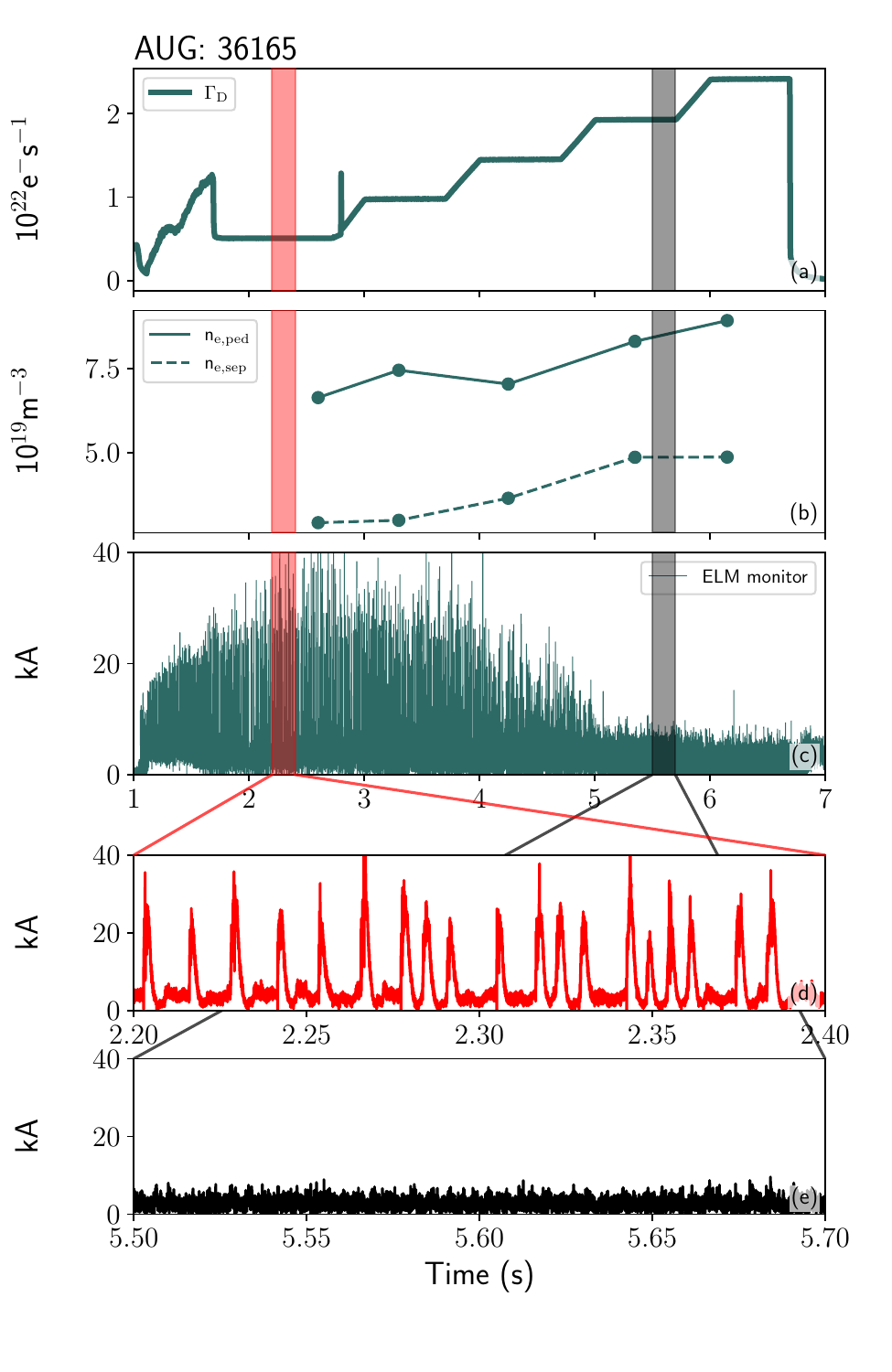}
    \caption{Plots of the (a) main ion fuelling, (b) pedestal top (solid) and separatrix (dashed) density and ELM monitor during the flat-top of an AUG QCE discharge. Middle: Zoom onto the ELM monitor during the ELMy phase of the discharge. Bottom: Zoom onto the ELM monitor during the QCE phase, highlighting the difference in amplitude of the QCE filaments in comparison to large Type-I ELMs.}
    \label{fig:qce_example}
\end{figure}

The quasi-continuous exhaust regime is a regime which, in contrast to NT, is operated at high positive plasma shaping. The regime is typically accessed at or close to the highest shaping possible in a particular device. A high separatrix density\cite{Harrer2018,Faitsch2023} has also been identified as a key entry requirement for the regime; on present-day devices this usually also leads to a high pedestal top collisionality owing to the coupling between separatrix and pedestal top. An example of a typical QCE discharge at ASDEX Upgrade is shown in figure \ref{fig:qce_example}.

The plasma shape is fully established at 2~s in this discharge and followed by a series of steps in the main ion fuelling, as shown in panel (a). This results in an increase of both the pedestal top and separatrix densities (panel b), with the separatrix density increasing by almost 40$\%$. A detailed view of the ELM signatures (panel c) from the poloidal current measured at the divertor tiles (which corresponds to the divertor heat loads\cite{Faitsch2023}) for an ELMy time window and a QCE time window are shown in panels (d) and (e), highlighting the substantial difference in heat loads from large ELMs and the QCE filaments. 

A highlight of the EUROfusion research program in recent years was the demonstration of the QCE at JET\cite{Faitsch2025,Faitsch2025a}, enabled by understanding of the operational space of the QCE from the separatrix parameters\cite{Harrer2018,Faitsch2023} and MHD stability\cite{Radovanovic2022,Dunne2024}; a sufficiently high shaping parameter (defined as $\mathrm{S_d}=\kappa^{2.2}(1+\delta)^{0.9}$) and a critical separatrix density which can be determined from machine parameters are the key requirements. 

For a given plasma shape and q$_{95}$ the critical $\alpha_\mathrm{edge}$ at the separatrix can be calculated using HELENA\cite{Huysmans1991}; this $\alpha_\mathrm{edge,crit}$ is a proxy for the threshold for KBMs, which are suspected to be the real limiting instability\cite{Snyder2009}. This is typically done assuming a single value of the separatrix density, consistent with the standard EPED assumption\cite{Snyder2009}. Some change of the critical $\alpha_\mathrm{edge}$ could be expected depending on the form of the density profile and its subsequent impact on the bootstrap current and magnetic shear. However, given that $\alpha_\mathrm{edge}$ is located close to the separatrix, this is not expected to have a significant impact on the results. The resulting $\alpha_\mathrm{edge,crit}$ for an AUG scenario is shown as the horizontal dashed purple line in figure \ref{fig:qce_os}.

\begin{figure}
\centering
\includegraphics[width=0.5\textwidth]{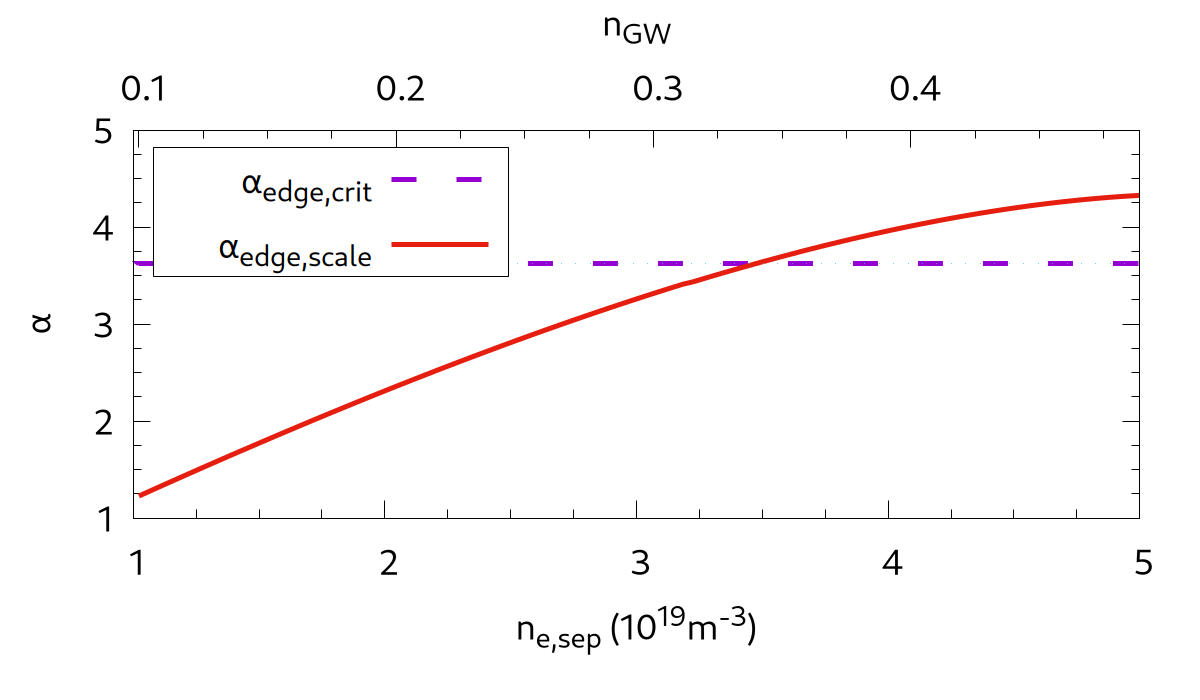}
    \caption{Critical (purple) and expected (red) $\alpha_\mathrm{edge}$ vs n$_\mathrm{e,edge}$. The critical separatrix density is found at the intersection of both lines, above which one expects access to the QCE regime.}
    \label{fig:qce_nesep}
\end{figure}

This critical $\alpha_\mathrm{edge}$ can be transformed into a critical n$_\mathrm{e,edge}$ as follows: an expected $\alpha_\mathrm{edge,exp}$ is constructed as $\mathrm{Rq^2}\beta/\lambda_\mathrm{p,scale}$ with $\beta\propto\mathrm{n_{e,edge}T_{e,edge}}$. T$_\mathrm{e,edge}$ is determined from the input heating power and Spitzer-Harm conduction, while $\mathrm{n_{e,edge}}$ is used as a free parameter to scan. The final variable is $\lambda_{p,scale}$ which can be taken from a scaling of AUG discharges\cite{Eich2020,Faitsch2023} and is given as $1.55(1+0.61\alpha_t^{2.35})\rho_s$, where $\alpha_t$ is a collisional broadening parameter and $\rho_s$ is the poloidal ion Larmor radius; all values are calculated self-consistently with the physical and magnetic geometry plus the power crossing the separatrix. This scaling for $\lambda_p$ 

When comparing the predicted $\alpha_\mathrm{edge,crit}$ from HELENA and the expected $\alpha_\mathrm{edge,exp}$ it should be noted that the definitions are slightly different; HELENA uses the Miller definition of $\alpha$\cite{Miller1998}, while the definition in the previous paragraph is a cylindrical approximation, with the simplification reflecting available experimental data.

This results in an expected $\alpha_\mathrm{edge}$ as shown by the red line in figure \ref{fig:qce_nesep} for an ASDEX Upgrade scenario. The minimum n$_\mathrm{e,sep}$ for the onset of separatrix ballooning modes in this scenario is predicted to be $0.33\mathrm{n_{GW}}$, with $\mathrm{n_{GW}} = \mathrm{I_p (MA)}/(\pi\mathrm{a}^2) (10^{20}\mathrm{m}^{-3})$. In the same way, predictions of the critical separatrix density can be made for any scenario on any device given only engineering parameters, assuming that the same scaling for $\lambda_p$ holds. 

\subsection{QCE access}
\begin{figure}
\centering
\includegraphics[width=0.5\textwidth]{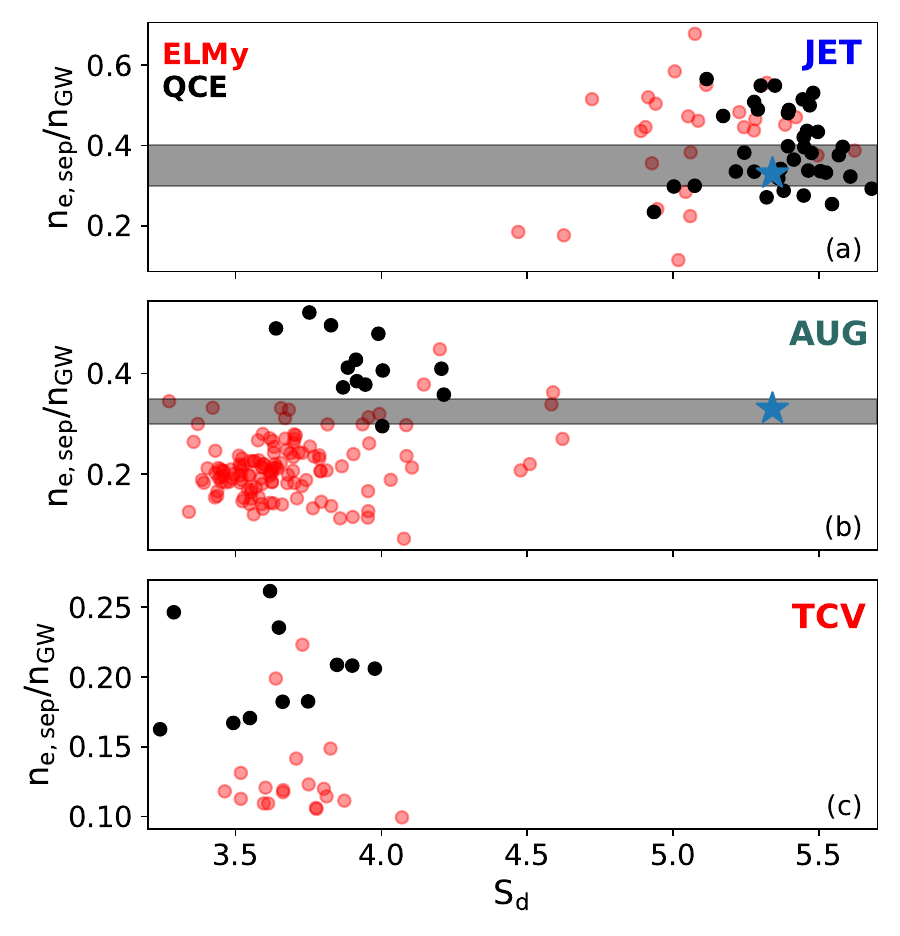}
    \caption{Normalised separatrix density vs plasma shaping for a range of scenarios in (a) JET, (b) AUG, and (c) TCV. Red points indicate data from ELMy plasmas, while black indicates QCE phases. The horizontal grey box indicates the range of minimum normalised separatrix density for QCE access, while the stars show the predicted ITER minimum normalised density.}
    \label{fig:qce_os}
\end{figure}

The model described in the previous section posits access to the QCE regime at high shaping and high separatrix density. To test this on the three devices, an operational space of normalised separatrix density vs shaping parameter is shown for JET, AUG, and TCV in figure \ref{fig:qce_os}. A range of plasma current and toroidal field with different q$_{95}$ values in each device are shown. ELMy data points are shown as the red points, and QCE data as black points. 

All of the experiments in JET shown here were performed with a high fuelling rate and, hence, the separatrix density does not vary significantly across the data set. Instead, the plasma shape is the main dividing factor between ELMy and QCE data. In AUG and TCV the more typical picture emerges with the QCE points existing at higher separatrix density. The grey boxes overlaying the JET and AUG data indicate the range of minimum predicted densities derived by the model from the previous section, which agrees well with the measured densities in both devices. Despite this, the model predicts separatrix densities far in excess of those observed in TCV; an intersection of $\alpha_\mathrm{edge,crit}$ with $\alpha_\mathrm{edge,scale}$ is only found by removing the $\alpha_t$ dependence of the gradient lengths, particularly $\lambda_p$, and lies at approximately 50$\%\mathrm{n_{GW}}$; one should note that there is no reasonable justification for this step, it is rather an attempt to quantify the departure from the agreement seen in the JET and AUG data sets. While the QCE does occur at higher density than the corresponding ELMy discharges, further investigations are necessary to explain the onset of QCE in TCV. In particular, the scaling used for $\lambda_p$ has large uncertainties, which impact all of the predictions shown here, and it is not necessarily given that such an extrapolation should apply to TCV.

Also overlayed on the AUG and JET data is a blue star indicating the predicted minimum separatrix Greenwald fraction for ITER at the ITER shape\cite{Dunne2024}. This shows that not only is the QCE predicted for ITER, but that the ITER shaping has already been tested at JET, and the normalised and absolute separatrix densities for QCE entry in ITER has already been achieved in AUG and JET. The QCE can therefore be expected to be accessible in ITER, as its entry requirements are met by the expected operating conditions of the device (medium-high separatrix density and high plasma shaping)\cite{Pshenov2025}.

\subsection{Pedestal performance}
\begin{figure}
\centering
    \includegraphics[width=0.5\textwidth]{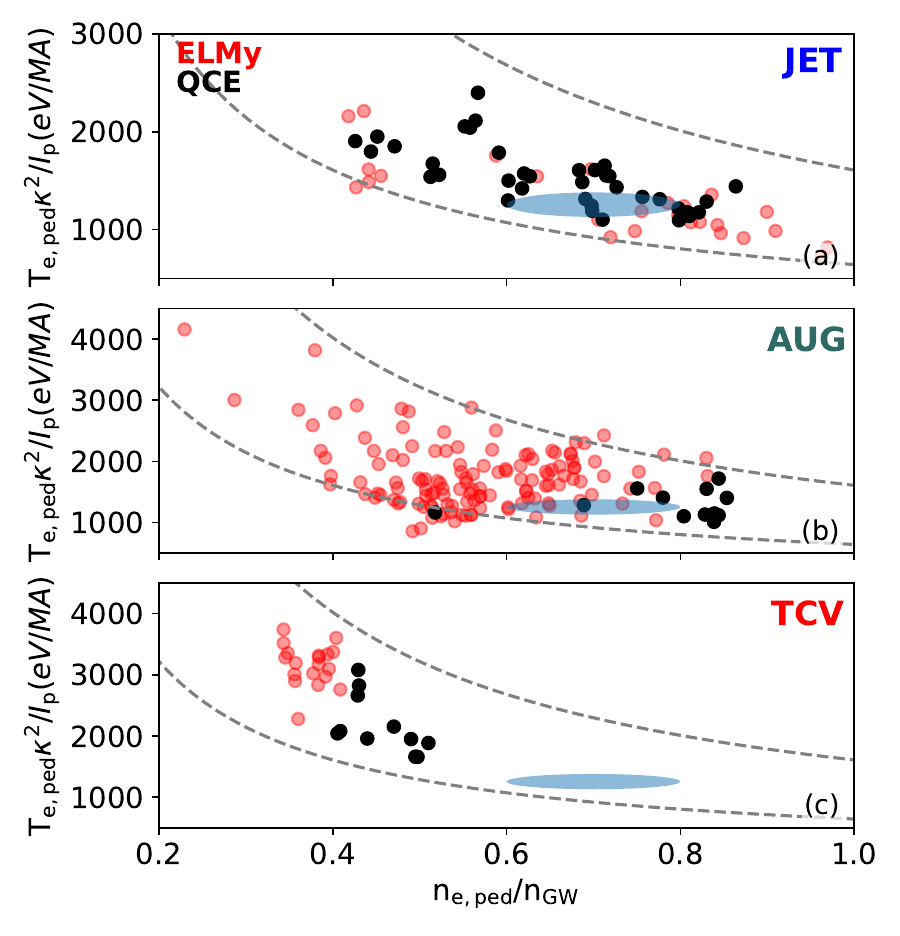}
    \caption{Normalised pedestal top temperature vs density for (a) JET, (b) AUG, and (c) TCV in QCE (black) and ELMy (red) plasmas. The isobars (dashed grey) indicate lines of constant $\beta_{e,pol,ped}$ with values of 0.08 and 0.2 for each device. The blue ellipises in each figure indicate the projected operational space of the ITER pedestal.}
    \label{fig:tene}
\end{figure}

Of particular interest for projection of the QCE regime to future devices are conditions at the pedestal top. With this in mind, figure \ref{fig:tene} compares the pedestal top temperature and density for JET, AUG, and TCV in QCE and ELMing plasmas for the same scenarios shown in figure \ref{fig:qce_os}.

Since a range of plasma currents, magnetic fields, plasma shapes, and q$_{95}$ values are considered, a fair comparison is best made by normalising the pedestal top values. A useful normalisation for the pressure is to represent it as $\beta_{pol,ped}$, as this is broadly representative of the drive for MHD instabilities in the pedestal\footnote{The normalised pressure gradient $\alpha$ can also be thought of as the gradient of the local $\beta_{pol}$ and the current density is dominated by the bootstrap current, which is also related to $\beta_{pol}$.}. The magnetic pressure is defined as B$_{pol}^2/(2\mu_0)$, with B$_{pol}\approx\mu_0\mathrm{I_p}/(2\pi a\kappa)$. 

A natural choice for the density is to normalise it to the Greenwald density n$_{GW} = \mathrm{I_p}/(\pi a^2)$. To keep the rest of the dependencies in the $\beta_{pol}$ definition, this implies a normalisation fo the temperature to $\mathrm{I_p}/\kappa^2$. This renders the ''isobars'' (dashed grey lines) in figure \ref{fig:tene} lines of constant $\beta_{e,pol,ped}$, the electron pedestal top poloidal $\beta$. The plot ordering and color scheme are the same as in figure \ref{fig:qce_os}. The dashed grey lines in all three subplots show the same $\beta_{e,pol,ped}$ values (0.08 and 0.2, accounting for the remaining factors of e$^{-1}, \mu_0$, and $\pi$ in the definition of $\beta_{pol}$), indicating that the same range of $\beta_{e,pol,ped}$ values are obtained in all three devices. 

With this normalisation, no difference between ELMy H-mode pedestal top values and those in the QCE regime can be discerned. In fact, as a result of how the experiments were performed, the JET data completely overlap in the two regimes, not only in $\beta$ but also in the normalised temperature and density. The AUG and TCV QCE plasmas exist, as might be expected, at higher normalised pedestal top density. The AUG data are quite beneficial in this comparison as they cover quite a wide range of plasmas, including those aimed at reaching low collisionality. With this in mind, it is interesting to note that the QCE still exists in the same $\beta_{e,pol,ped}$ range. 

Finally, the blue ellipses overlayed on the three plots indicate a range of expectations for the ITER 15 MA baseline pedestal, with a Greenwald fraction between 0.6-0.8, and an expected pedestal top temperature of 4-5.5~keV, again demonstrating the compatibility of the ITER pedestal with the QCE regime. The ASDEX Upgrade and JET data overlap in both normalised temperature and density, while the TCV data exist at lower normalised density. The compatibility of ITER and high pedestal performance in conjunction with the existence of edge ballooning modes has also been shown in several other works\cite{Maget2013,Radovanovic2022,Luda2025}. 

\subsection{The QCE in DT at JET}
\begin{figure}
\centering
\includegraphics[width=0.5\textwidth]{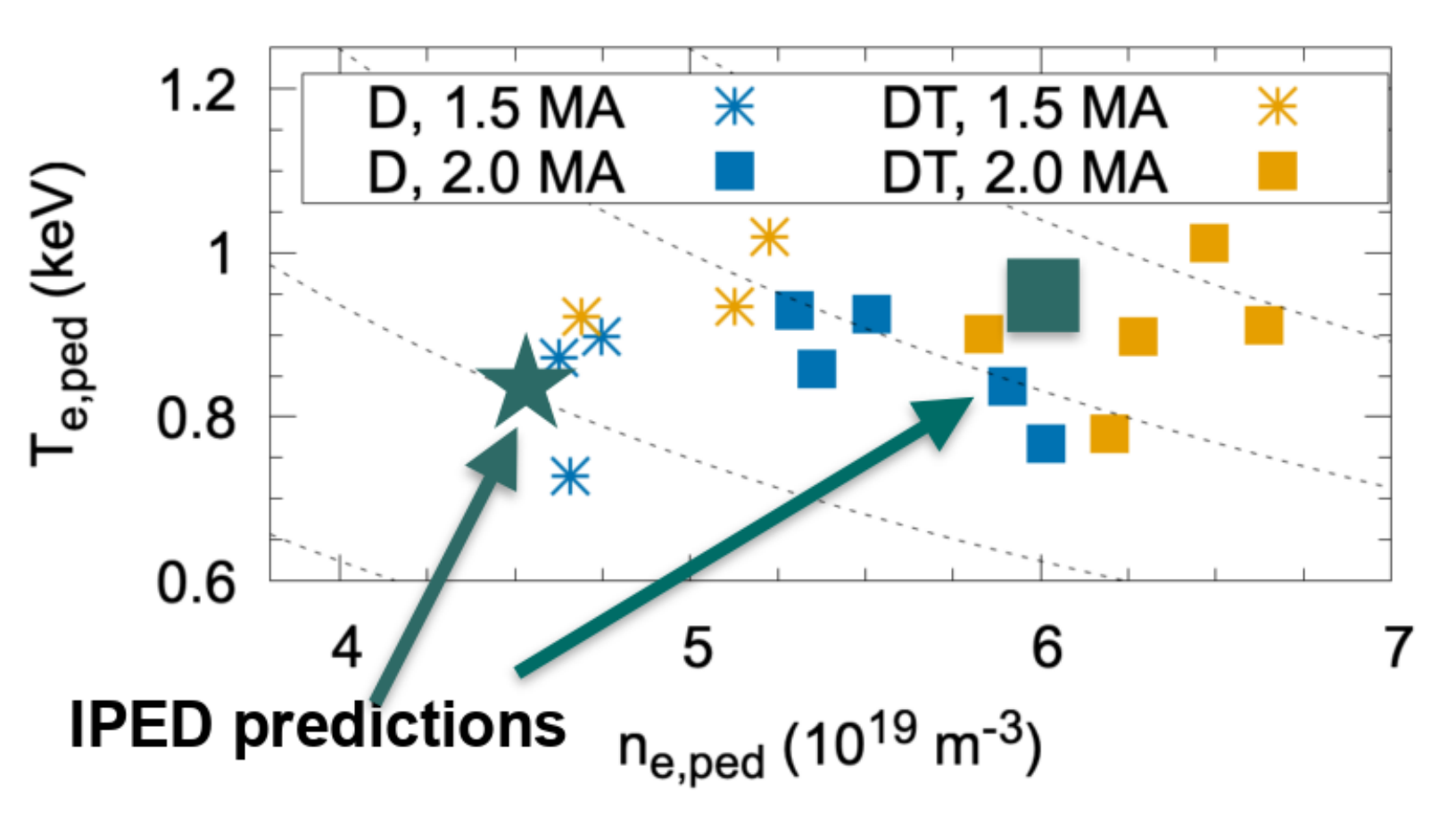}
    \caption{Temperature vs density for two JET QCE scenarios at 1.5 MA (stars) and 2 MA (squares) in D (blue) and DT (gold) operation. The green shapes correspond to IPED predictions for the pedestal height in both scenarios.}
    \label{fig:jetped_dt}
\end{figure}

The successful demonstration of the QCE in DT at JET has already been reported\cite{Faitsch2025,Kappatou2025}, highlighting the easy transfer of the scenario from D to DT operation. The confinement improvement, often observed in ELMy H-modes when switching to DT operation was also observed in the QCE. Figure \ref{fig:jetped_dt} shows the pedestal top temperature and density (both without normalisation) for two JET QCE scenarios at 1.5~MA (stars) and 2.0~MA (squares), each with a q$_{95}$ of 4.0, for D (blue) and DT (gold) plasmas. The increase of pedestal density at constant temperature with increasing plasma current is clear. Also clear is the increase in pedestal top pressure, mainly coming from an increase in density when changing from D to DT plasmas, as has been already reported\cite{Faitsch2025}. 

Overlayed on figure \ref{fig:jetped_dt} are green shapes corresponding to IPED predictions for the pedestal height in each of the scenarios. These predictions were made with the "standard" width pre-factor of 0.076\cite{Snyder2009}, and so represent a conservative estimate for the pedestal height while underestimating the pedestal width. The other inputs are the actual plasma shape used, I$_\mathrm{p}$, B$_\mathrm{T}$, the pedestal top density (matching the locations in the figure), and the global $\beta_\mathrm{N}$ corresponding to the DT pulses). Nevertheless, this indicates that the pedestal in QCE plasmas is broadly consistent with the predictions for an ELMy H-mode pedestal.

\subsection{Extrapolation to ITER}
The ability to predict the minimum separatrix density and expected pedestal top density in the QCE regime using ideal MHD lends further support to the QCE as a plasma scenario for ITER. Previous work\cite{Maget2013,Radovanovic2022,Luda2025} has indicated that ballooning modes are expected in the pedestal of  the 15~MA baseline scenario, while maintaining high pedestal top temperature and pressure. Recent work\cite{Dunne2024} has also shown that the minimum separatrix density required for QCE access in ITER is approximately 30$\%\mathrm{n_{GW}}$, which is in the range of densities achieved in present-day machines. While the normalised global confinement time in QCE scenarios is often below 1.0, the pedestal performs as expected by ideal MHD, meaning that an integrated modelling code, such as IMEP\cite{Luda2025}, can make better predictions for the global confinement when extrapolating a high density scenario, such as the QCE.

\section{Comparison of NT and QCE at JET}\label{sec:qce_nt}
\begin{figure}
    \centering
    \includegraphics[width=0.5\textwidth]{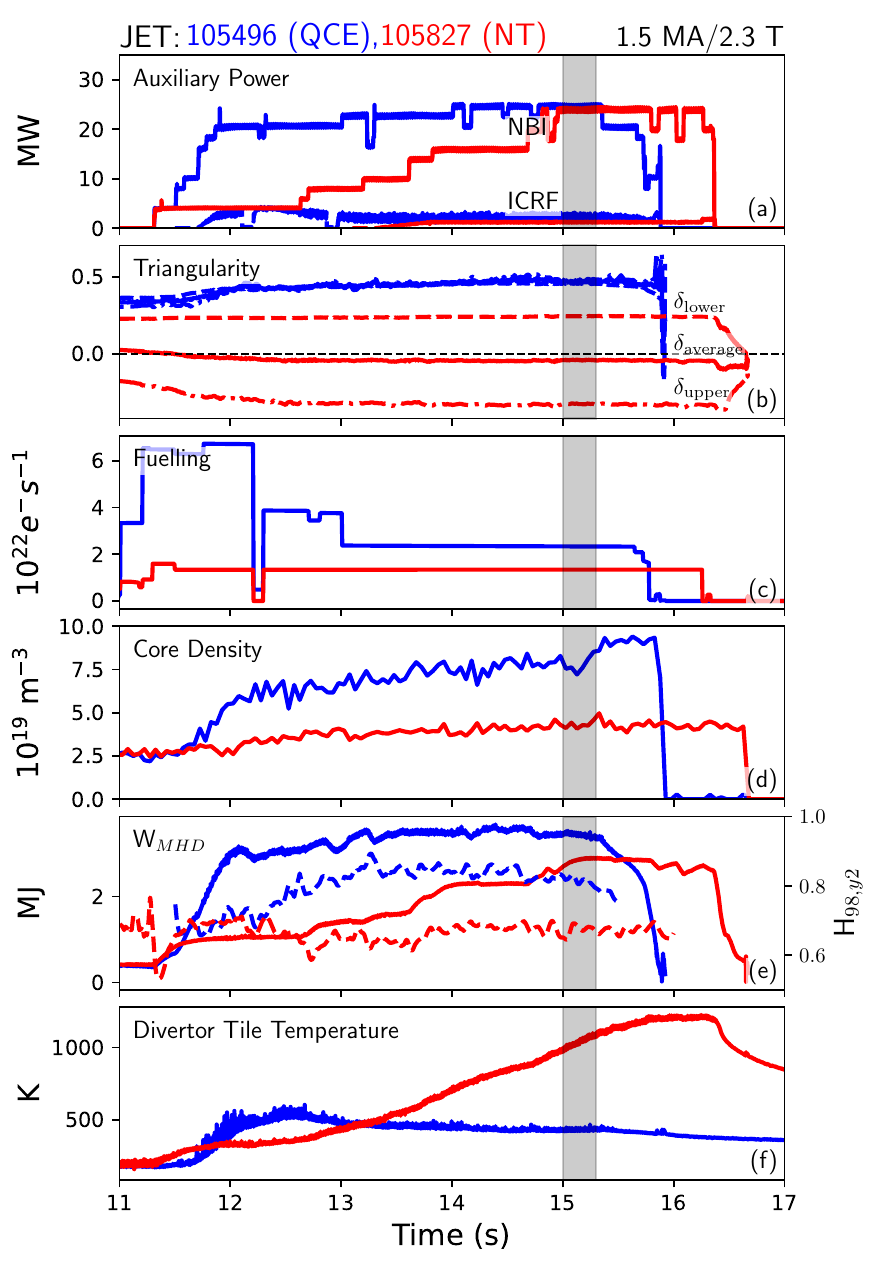}
    \caption{Time-traces of (a) plasma heating power, (b) upper, lower, and average triangularity, (c) D fuelling rate, (d) core density measured by Thomson Scattering, (e) plasma stored energy (solid) and normalised confinement time (dashed), and (f) tile temperature for JET QCE (blue) and NT (red) discharges.}
    \label{fig:jet_nt_qce}
\end{figure}

One of the goals of the negative triangularity experiments at JET was to achieve a scenario comparable to the QCE, i.e. at 1.5~MA and 2.3~T with similar heating power. This has enabled the first direct comparison of these regimes in terms of both plasma performance and ELM-avoidance. Time-traces of the two scenarios are shown in figure \ref{fig:jet_nt_qce}. 
Panel (a) shows the heating power for the QCE (blue) and NT (red) discharges, panel (b) the core plasma density, (c) the plasma stored energy (solid) and H$_{98,y2}$ (dashed, right axis), (d) the W-I ELM monitor signal, and (e) the tile-temperatures on the horizontal divertor target for the QCE discharge (the strike-points in the NT discharge did not allow tile-temperature measurements). 

There are several striking differences between the discharges, notably in the plasma density, which is significantly higher in the QCE discharge; this is to be expected from the difference in the shapes and expected particle confinement improvement at high positive triangularity. While the normalised plasma confinement H$_{98,y2}$ is significantly lower in the NT discharge (0.6 vs 0.85), the higher volume of the NT shape leads to the plasma stored energy being quite comparable between the two discharges. The ELM-monitor also shows a significant difference between the two discharges; despite both scenarios being limited by ballooning modes, the QCE shows strong filamentary activity throughout the pulse. Despite this, the tile temperature measurements shown only a small excursion, as has been previously shown\cite{Faitsch2025}. In contrast, the NT discharge only begins to show comparable filamentary activity later in the discharge at the highest heating power.

\begin{figure}
    \centering
    \includegraphics[width=0.5\textwidth]{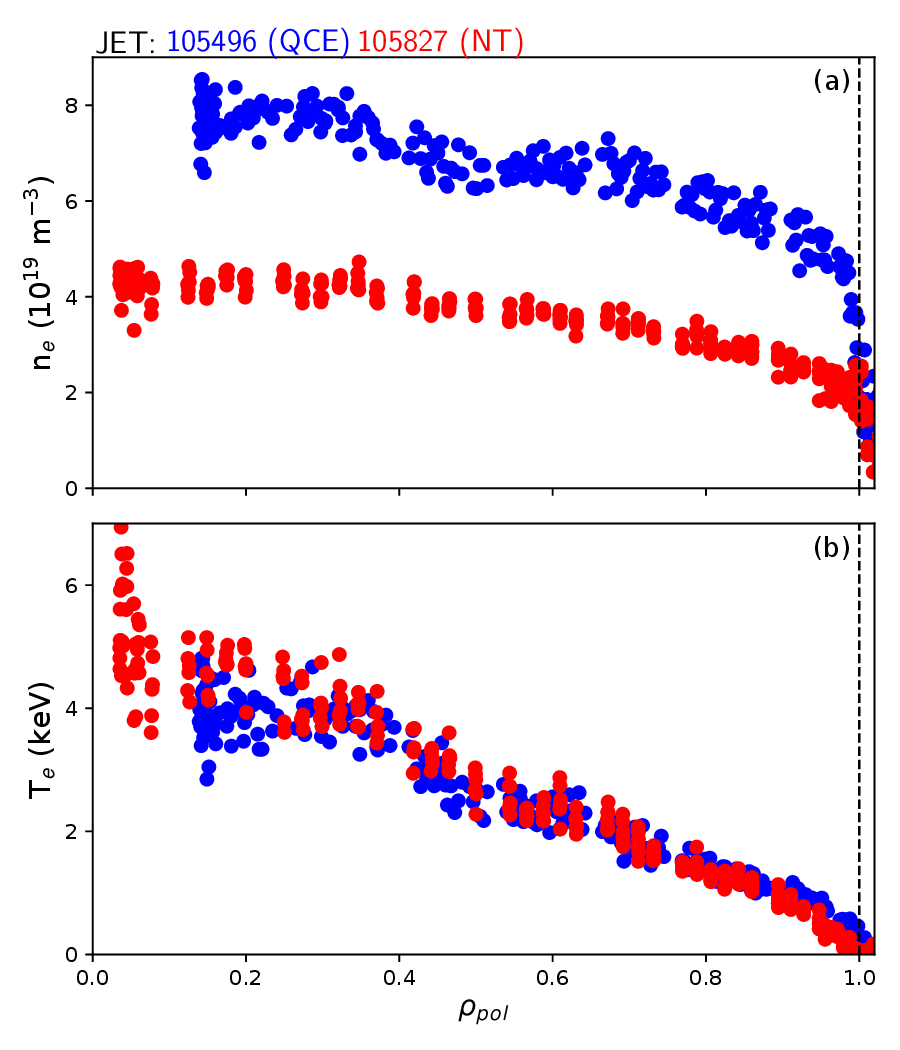}
    \caption{Profiles of electron density (a) and temperature (b) for JET QCE (blue, $\#105496$) and NT (red, $\#105827$) discharges (data are taken from both pulses between 55.0-55.3~s); the scenarios exhibit similar electron (and ion, not shown) temperature profiles, but the QCE discharge has a significantly higher density, stemming from a strong pedestal gradient.}
    \label{fig:jet_nt_qce_profiles}
\end{figure}

A comparison of the electron temperature and density profiles for the two discharges is shown in figure \ref{fig:jet_nt_qce_profiles}. The electron temperature is almost identical over the entire plasma radius (T$_\mathrm{i}$ profiles, which are not shown here, are also similar), while the density, as expected from the time-traces, is significantly higher in the QCE discharge. This difference stems from the pedestal, which exhibits a strong density gradient in the QCE discharge. In the NT discharge, the density gradient in the edge region is comparable to that in the plasma core; this may be due to expected high ballooning transport across the entire pedestal region in the NT shape and is being investigated.

This comparison highlights the relative strengths of the two approaches to ELM-avoidance. The QCE regime builds on decades of experience with standard H-mode-like plasmas, using high positive triangularity to maximise the plasma confinement at high density, which can be expected to extrapolate to future machines in a manner given by multi-machine confinement time scalings. NT plasmas, on the other hand, are a more novel approach, avoiding the H-mode entirely and leveraging a combination of the high plasma volume (the nature of the shape places more plasma at higher major radius than a positive triangularity counterpart) and, depending on the shape, also improved confinement radially inwards of the plasma edge to recover H-mode-like confinement. Filaments are often associated with the QCE, and an open research question is how these filaments will scale to reactors, particularly the energy deposited not only on the divertor but also the first wall. NT plasmas do not yet appear to show filaments to the same extent, though the area of divertor and first-wall power loads is still an open investigation for NT scenarions. 

A general conclusion cannot yet be drawn on the efficacy of either scenario for a potential reactor. However, the advances in understanding and predictive modelling for both QCE and NT plasmas has enabled us to make predictions for future machines in terms of critical shaping and, in the case of the QCE, the minimum separatrix density for ELM-avoidance. This allows an evaluation for any planned devices to be made for these two scenarios to enable further tests of these models to form a more complete physics basis for an eventual reactor plasma, including global performance. 

\section{Summary and conclusions}
The work presented here has shown the current understanding of the physics mechanisms for access to the QCE and NT large ELM-avoidance regimes. In both cases, ballooning modes play a dominant role. For the QCE, a ballooning mode at the separatrix (modelled using an ideal ballooning mode and in reality, likely a KBM of some kind), while for NT blocking access to second stability in the middle of the pedestal is a requirement for avoiding large ELMs. The understanding gained from experiments across both EUROfusion and other international devices allows predictions to be made, making scenario development on devices significantly easier. 

The stepladder approach within EUROfusion, understanding physics on smaller devices before applying it to larger ones, notably JET, has significantly shortened scenario development time necessary for both QCE and NT demonstration in JET; without guidance from modelling and experimental validation on AUG and TCV, neither the QCE nor NT would have been possible in the experimental time available. The successful demonstration of the QCE in D plasmas in JET also paved the way for a successful series of QCE experiments in DT plasmas. This important step highlights that the lessons learned during non-nuclear phases can be transferred to DT operation in future devices. Predictions for the minimum density for QCE access in ITER indicate that the regime should be accessible in ITER, and pedestal predictions made with ideal MHD codes are also expected to apply in the QCE regime in ITER. SPARC and EU-DEMO-like tokamaks are also expected to easily access the QCE regime owing to the large poloidal field in these devices; the high shaping required for QCE, which is planned for these devices also pairs well with performance optimisation requirements to maximise fusion power. %In both SPARC and EU-DEMO, 

NT ELM avoidance has been successfully demonstrated on both AUG and JET, guided by modelling and similarity experiments in TCV. Since the minimum shaping required can be predicted by ideal MHD, the required triangularity for any future devices can also be predicted, and tested on currently running machines to further validate the model. In addition, while lower divertor heat loads are expected for NT plasmas due to the larger major radius of the strikepoints, this must still be verified experimentally.

Work is still ongoing in the EUROfusion program to extend not only the QCE and NT regimes to other devices, but also the EDA H-mode, I-mode, MP ELM-suppression, and QH-mode. The expansion of the QCE operational range to higher current and even lower q$_{95} (<3.5)$ than has already been achieved\cite{Faitsch2025} has a high priority. While it has been shown for a single plasma\cite{Faitsch2023}, robust ELM avoidance during the current ramp and across the LH transition is also a high priority for future experiments; with guidance from modelling, significant progress in feedback control of such experiments is expected. Finally, the plasma-wall interaction in QCE plasmas must also be understood, as the fall-off lengths in the SOL become significantly longer than in ELMy H-mode plasmas\cite{Redl2024, Sun2025} and may increase the first-wall loads.

The final test of the QCE regime will come with the next generation of devices which are currently being commissioned and built; JT-60SA, SPARC, and DTT. Thes latter two devices devices will allow a more complete test of core-edge integration than is possible with current devices, in particular the combination of the high separatrix density and collisionality required for power exhaust with the low pedestal top collisionality expected in high fusion gain scenarios. Modelling has allowed predictions to be made for future devices, and the validation efforts taking place on present-day machines will allow targeted scenario development, as was displayed during the JET experiments. Combined with integrated models, the physics of ELM-free regimes can be better projected to future devices, ensuring safe machine operation, while fulfilling their fusion power mission.

\section*{\hfill \textbf{ACKNOWLEDGEMENTS} \hfill}
This work has been carried out within the framework of the EUROfusion Consortium, partially funded by the European Union via the Euratom Research and Training Programme (Grant Agreement No 101052200 — EUROfusion). The Swiss contribution to this work has been funded by the Swiss State Secretariat for Education, Research and Innovation (SERI). Views and opinions expressed are however those of the author(s) only and do not  necessarily reflect those of the European Union, the European Commission or SERI. Neither the European Union nor the European Commission nor SERI can be held responsible for them. This work was also supported by the US Department of Energy under Grants DE-SC0014264.

\section*{References}

\bibliographystyle{new}
\bibliography{extracted}

\end{document}